\documentstyle[aps,epsfig]{revtex}

\newcommand{\ket}[1]{\mbox{$ \mid\!{#1}\,\rangle $}}
\newcommand{\bra}[1]{\mbox{$ \langle\,{#1}\!\mid $}}

\begin{document}
\title{Josephson spectroscopy of a dilute Bose-Einstein condensate in a
double-well potential}
\author{E. Sakellari, M. Leadbeater, N. J. Kylstra, and C. S. Adams}
\address{Department of Physics, University of Durham, 
Durham DH1 3LE, United Kingdom}
\date{\today}

\draft
\twocolumn
\maketitle

\begin{abstract}

The dynamics of a Bose-Einstein condensate in a double-well potential
are analysed in terms of transitions between energy eigenstates. 
By solving the time-dependent and time-independent Gross-Pitaevskii
equation in one dimension, we identify tunnelling resonances 
associated with level crossings,
and determine the critical velocity that characterises the
resonance. We test the validity of a non-linear two-state model, 
and show that for the experimentally interesting
case, where the critical velocity is large, 
the influence of higher-lying states is important. 

\end{abstract}

\pacs{03.75.Fi}

\section{Introduction}

The behaviour of dilute Bose-Einstein condensates in multi-well
potentials is remarkably rich. The tunnelling of a condensate through
an optical lattice potential \cite{ande98,cata01} provides an atomic
physics analogue of a Josephson junction array. In principle, the
analogue of a single Josephson junction can be realised experimentally
by a condensate in a double-well potential \cite{ande97}, and this
system has attracted considerable theoretical interest
\cite{milb97,angl97,smer97,ruos98,ragh99,giov00,will01,meie01}. However, the 
experimental observation of the transition between the dc and ac
Josephson effects is challenging because the
small energy splitting associated with Josephson oscillations means
that thermal or quantum fluctuations will tend to destroy the effect 
even at the lowest achievable temperatures 
\cite{angl97,ruos98,meie01,vard01,pita01}. While the energy splitting can be increased
by lowering the barrier height, it then becomes comparable
with that of higher-lying states, and the accuracy of a
two-mode approximation to describe the system \cite{milb97,smer97,ragh99,will01} becomes
questionable. 

In this paper, we investigate the Josephson dynamics of a dilute 
Bose-Einstein condensate in a double-well potential, by considering
transitions between energy eigenstates.
We show that in the regime of
most interest for experiments where the energy splittings are large, 
the influence of higher-lying states cannot be ignored.
The paper is organised as follows.  First, we calculate
the eigenenergies of the double-well potential as a
function of the barrier position. We show that in the
vicinity of a level crossing the non-linearity leads
to triangular structures in the eigenenergy curves. Similar 
structures have been shown to occur in a non-linear Landau-Zener model 
\cite{wu00}, and near the zone boundary in optical lattices \cite{wu01,diak01}.  
Next we review the two-state model \cite{milb97,smer97,wu00}, 
and show that as long as the influence
of higher-lying states is small, the level structure 
can be reproduced using
three parameters determined by numerical
solution of the time-independent Gross-Pitaevskii equation. Finally in Section \ref{sec:dynamics} we consider the 
time-dependent evolution as the barrier is moved, 
and analyse the transition
between dc and ac tunnelling in terms of transitions
between eigenstates. 

\section{Eigenenergies of the double-well potential}
\label{sec:newton}

The eigenenergies of the double-well potential, $V_{x_0}$, are found by
numerical solution of the one-dimensional (1D) Gross-Pitaevskii (GP) equation
\begin{equation}
i \partial_t \psi(t)= -\textstyle{1\over2}\nabla^2\psi(t) +
               V_{x_0}\psi(t)+g\vert\psi(t)\vert ^2\psi(t)~.
\label{eqn:GPE}
\end{equation}
Here $g=4\pi na$ is the self interaction parameter, 
$a$ is the $s$-wave scattering length, and 
$n$ is the number of atoms per unit area in the $y-z$ plane. 
Distance, time and energy are measured in terms of the 
standard harmonic oscillator units
(h.o.u.), $( \hbar /m \omega)^{1/2}$,
$\omega^{-1}$ and $\hbar \omega$, respectively.
In particular, we have concentrated on a system confined
by the potential
\begin{eqnarray}
V_{x_0}=\textstyle{1\over 2}x^2+h~{\rm exp}[-(x-x_0)^2]~,
\end{eqnarray}
describing a harmonic trap with a Gaussian potential barrier with
height $h$ and unit width. We have also carried out three-dimensional
simulations, and verified that the essential physics is described 
by a 1D model. This model facilitates 
an exploration of a wide range of values of the self-interaction, 
$g$, barrier height, $h$, and
barrier velocity.

Time-independent states of the form,
$\psi(x,t)={\rm e}^{-i\mu t}\varphi(x)$, where $\mu$ is the 
chemical potential,
are found by numerical solution of the discretized system of equations,
\begin{eqnarray}
\label{eq:nlin}
f_{i} & \equiv & -\left(\varphi_{i+1}+\varphi_{i-1}
-2 \varphi_{i} \right)/2 \Delta^2
+  V_{i} \varphi_{i}+  g\varphi_{i}^3
-  \mu \varphi_{i}\nonumber\\
& = & 0~,
\label{eq:dgp}
\end{eqnarray}
where $\varphi_{i}=\varphi(x_i)$ is a real function,
and $\Delta=1/75$ is the grid spacing.
Note that in general, one must solve for both the real and imaginary parts of
$\varphi(x_i)$ \cite{wini99}. 
Since $\mu$ is also a variable we have
$n$ equations and $n+1$ unknowns.
An additional equation is provided by fixing the normalisation of the wave function,
\begin{equation}
f_{n+1}  \equiv  \sum_{i=1}^n \varphi_i^2 -1 =0~.
\label{eq:norm}
\end{equation}
A solution is found iteratively by linearising equations (\ref{eq:dgp}) 
and (\ref{eq:norm}) (with $\mu\equiv \varphi_{n+1}$)
\begin{equation}
f_{i} ( \varphi_{i}^{(p)}) + \sum_{i=1}^{n+1}
 \left( \varphi_{i}^{(p+1)} - \varphi_{i}^{(p)} \right)
\left[ {\partial f_{i} \over  {\partial \varphi_{i}}} \right]^{(p)}
\approx 0~,
\label{eq:lin}
\end{equation}
where $\varphi^{(p+1)}$ is determined from the approximation $\varphi^{(p)}$
by
solving (\ref{eq:lin}) using the bi-conjugate gradient method
\cite{numrec}.
The iterative solution depends
on the symmetry of the initial trial wave function $\varphi^{(0)}$.
For example, to find the lowest energy excited state we begin 
with a ground state solution
at $x_0=0$ and change the parity. Higher lying excited states can be 
found by beginning
with an excited state solution of the harmonic trap and
moving the barrier in from the side.

\begin{figure}[hbt]
\centering
\epsfig{file=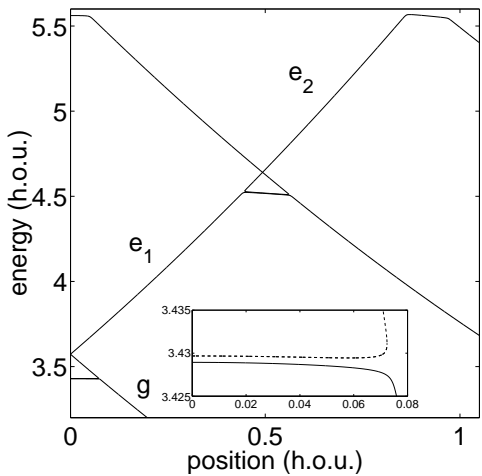,width=6.5cm,angle=0,clip=,bbllx=65,bblly=180,bburx=570,bbury=675}
\caption{The energy of states ground states $g$, and the 
first two excited states, $e_1$ and 
$e_2$ as a function of the barrier position 
with $h=12$ and $g=0.5$. In the vicinity of the 
level crossing the upper level is split into 
three states that form a triangular structure.
The ground-excited state splitting is shown inset.}
\label{fig:eh12}
\end{figure}

In Fig.~\ref{fig:eh12} we show the eigenenergies of the ground level $g$ 
and the first and second excited levels, $e_1$ and $e_2$, as a 
function of the barrier position $x_0$ with $g=0.5$ and $h=12$. 
A triangular structure appears in the upper level at each
level crossing. For states $g$ and $e_1$ this structure is 
essentially the same as that discussed by Wu and Niu in the context of 
a non-linear two-state model \cite{wu00}. The three states in the 
triangular region correspond to an
anti-symmetric state with the population balanced between
the two wells, and two higher energy states 
with most of the population 
in either the lower or the upper well. The higher energy states
are referred to as self-trapping states \cite{milb97,smer97}. Any
dissipative mechanism that damps the system towards the ground state
will tend to destroy self-trapping, as pointed out in Refs. \cite{angl97,ruos98,meie01}.

For small $x_0$ the energy of the ground state is almost independent of
the barrier position. This is because the transfer of atoms through the
barrier exactly compensates for the change in potential energy.
However, this cannot continue when all the atoms reach one
side and the energy becomes strongly position dependent again at the critical
displacement, $x_{\rm c}$ ($x_{\rm c}=0.07$ for the parameters in
Fig.~1). 

At $x_0\sim0.5$ there is a second level crossing
where it becomes energetically favourable for the first excited state
$e_1$ to move from the upper to the lower well. The energy level
structure is similar to that at $x_0=0$, where it is the ground state
population that moves from the upper to the lower well. In both cases,
following the eigenenergy curve produces a large change
or `step' in the well populations. When
considering a moving barrier these steps give rise to
a resonance in the tunnelling current. In Section \ref{sec:dynamics}, we suggest 
the optimal scheme for observing such tunnelling resonances.

\begin{figure}[hbt]
\centering
\epsfig{file=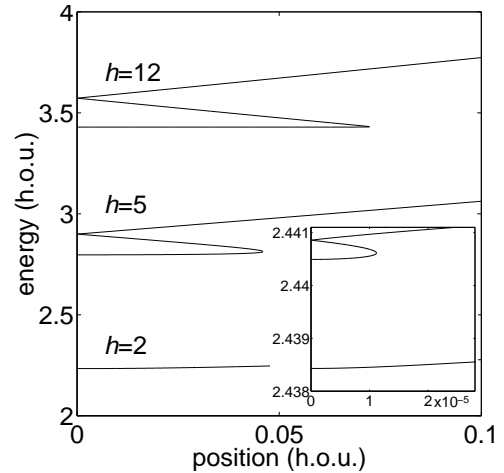,width=6.5cm,angle=0,clip=,bbllx=65,bblly=180,bburx=570,bbury=680}
\caption{
The energy of state $e_1$ as a function of the barrier position for $g=0.5$ and
barrier heights $h=2$, 5, and 12. The energy curves for $h=2.91$ and $h=2.92$ are shown
inset. The loop structure appears at $h=2.916$.}
\label{fig:e1hdep}
\end{figure}

The energy splitting between the ground and lowest excited state 
is extremely small for $h=12$, see Fig.~\ref{fig:eh12}(inset). 
From an experimental viewpoint, lower barriers resulting in
larger energy splittings are more interesting. 
In Fig.~\ref{fig:e1hdep} we illustrate how the triangular structures evolve as
a function of barrier height with $g=0.5$. 
The appearance of the loop structure, which coincides
with the threshold for self-trapping,
occurs at critical height $h=2.916$ (see Section~\ref{sec:eigen}). 
The structure becomes more
triangular as $h$ increases.

\begin{figure}[hbt]
\centering
\epsfig{file=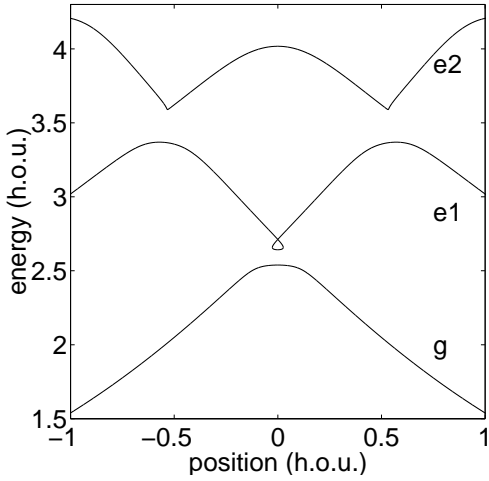,width=6.5cm,angle=0,clip=,bbllx=85,bblly=170,bburx=500,bbury=585}
\caption{The energy of states $g$, $e_1$ and 
$e_2$ as a function of the barrier position for $h=4$ and
$g=0.5$.}
\label{fig:eh4cp5}
\end{figure}

In Fig.~\ref{fig:eh4cp5} we show the eigenenergies for
$g=0.5$ and $h=4$. In this case, the loop structure appears for
the first but not the second excited state. The appearance of a 
loop structure results in the breakdown of adiabatic following
of an eigenenergy curve \cite{wu00}. In Section~\ref{sec:dynamics},
we will see that this breakdown is associated with a discontinuity
in the population difference between the two wells. 

\begin{figure}[hbt]
\centering
\epsfig{file=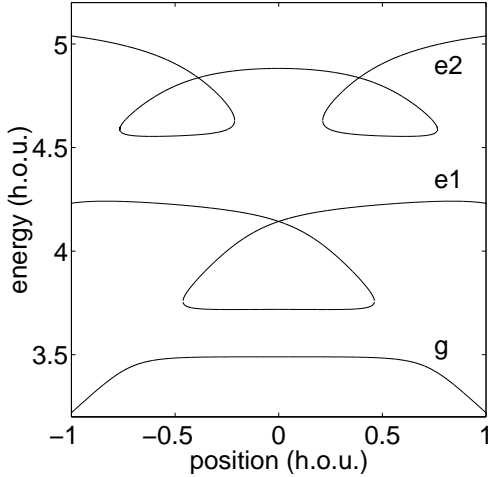,width=6.5cm,angle=0,clip=,bbllx=80,bblly=170,bburx=495,bbury=585}
\caption{The energy of states $g$, $e_1$ and 
$e_2$ as a function of the barrier position for a barrier height
$h=4$ and $g=5$.}
\label{fig:eh4c5}
\end{figure}

Finally in Fig.~\ref{fig:eh4c5}, we illustrate the 
effect of increasing the self-interaction parameter to $g=5$. 
In this case the energy splittings are a significant fraction
of the harmonic oscillator energy level spacing and the 
critical displacement coincides with
the position of the crossings between states $e_1$ and $e_2$.
In this regime, the influence of  state $e_2$
cannot be neglected, and a two-state approximation 
cannot be used to describe the system.

We stress that by solving the
Gross-Pitaevskii equation in three dimensions we have obtained
eigenenergy curves qualitatively similar to those shown in Figs.~\ref{fig:eh12} to \ref{fig:eh4c5}.

\section{Non-linear two-state model}
\label{sec:2mode}

In this section we review the two-state model \cite{milb97,smer97}, 
and apply it to the specific case of a
Bose-Einstein condensate in an asymmetric double-well. 
The wave function, $\psi(t)$, is written as
\begin{eqnarray}
\psi(t) &=& \phi_1(t)\psi_1 + \phi_2(t)\psi_2~,
\end{eqnarray}
where the zeroth-order (real) mode functions $\psi_i$, $i=1,2$, are
eigenfunctions of the time-independent GP equations
\begin{eqnarray}
\mu \psi_i  &=& 
-\textstyle{1\over2}\nabla^2\psi_i +V_i\psi_i+g\vert\psi_i\vert ^2\psi_i~.
\end{eqnarray} 
The mode potentials $V_1$ and $V_2$ are single-well potentials 
displaced to the left and the right of $x=0$, respectively, 
and satisfy $V_1(x)=V_2(-x)$ so that $\psi_1(x)=\psi_2(-x)$.
%In principle, the potentials would be chosen such that they 
%reproduce as closely as possible the ....
The time-dependent coefficients $\phi_i(t)$ can, in turn, be expressed
as
\begin{equation}
\phi_i(t)=\sqrt{N_i(t)}{\rm e}^{i\theta_i(t)}~,
\label{eqn:coeff}
\end{equation}
where $N_i(t)$ and $\theta_i(t)$ are the population and phase of state $i$,
with $N_1(t)+N_2(t)=1$.

Recalling that the two mode functions satisfy
$\bra{\psi_1}V_{x_0=0}\ket{\psi_1}=
\bra{\psi_2}V_{x_0=0}\ket{\psi_2}$
and redefining the zero of energy, we obtain the following non-linear
equations of motion for $\phi(t)=(\phi_1(t),\phi_2(t))^{\rm T}$:
\begin{eqnarray}
i\dot{\phi}(t)&=& H(\phi(t))\phi(t)~. 
\label{eqn:TDGPE}
\end{eqnarray}
The non-linear Hamiltonian matrix, $H(\phi(t))$, is given by
\begin{equation}
H(\phi(t))=\frac{1}{2} \left[\begin{array}{cc}
-\Delta + NE_{\rm C} & -E_{\rm J}\\
-E_{\rm J} &  \Delta - NE_{\rm C}
\end{array}\right]
\label{eqn:NLHam}
\end{equation} 
where $N=N_1-N_2$ is the population difference between the left
and right sides of the well, and $E_{\rm J}$ and $E_{\rm C}$ are the
coupling or Josephson energy  and the self-interaction energy, respectively. 
The self-interaction energy of each mode is
\begin{eqnarray}
E_{\rm C}= g \bra{\psi_1}\,\vert\psi_1\vert^2\ket{\psi_1}=
g \bra{\psi_2}\,\vert\psi_2\vert^2\ket{\psi_2}~,
\end{eqnarray}
and we define the shift in energy of the mode states due to the
displacement of the potential barrier to be
\begin{eqnarray}
\textstyle{1\over2}
\Delta &=&
   \bra{\psi_1}\left( V_{x_0=0}-V_{x_0}\right)\ket{\psi_1} \nonumber \\
       &=&
   \bra{\psi_2}\left( V_{x_0}-V_{x_0=0}\right)\ket{\psi_2}~.
\end{eqnarray}
If states $\psi_1$ and $\psi_2$ correspond to the population on the
left and right of the barrier, respectively, then linearising yields
\begin{eqnarray}
\textstyle{1\over2}
\Delta &\simeq& x_0
\bra{\psi_1}\left(\frac{dV_{x_0}}{dx_0}\right)_{x_0=0}\ket{\psi_1}
=-\alpha x_0~,
\end{eqnarray}
where $\alpha<0$ is the rate of change of the energy with
displacement.  As will be discussed below, this linear approximation
works well when the influence of higher-lying states
is negligible. Finally, the coupling energy of the two-modes
is
\begin{eqnarray}
\textstyle{1\over2}
E_{\rm J} &\simeq& -\bra{\psi_1} 
\left(-\textstyle{1\over2}\nabla^2 +V_{x_0=0}\right)\ket{\psi_2} ~,
\end{eqnarray}
where use has been made of the fact that
\begin{eqnarray}
\bra{\psi_1} \left(\frac{dV_{x_0}}{dx_0}\right)_{x_0=0}\ket{\psi_2} &=& 0~.
\end{eqnarray}
The Hamiltonian, equation (\ref{eqn:NLHam}),
is similar to that considered by Wu and Niu in their study of non-linear 
Landau-Zener tunnelling \cite{wu00}.
In the next section we discuss the eigenstates of $H$
in relation to the numerical solutions discussed in Section~\ref{sec:newton}.

\section{Eigenenergies of the two-state model}
\label{sec:eigen}

Within the two-state model, one may find the eigenenergies of the
system by substituting $\theta_1(t)=\theta_2(t)=\epsilon t$.  The
eigenenergies are given by the roots of the quartic equation
\cite{wu00}
\begin{eqnarray}
\epsilon^4-E_{\rm C}\epsilon^3 &+&
\textstyle{1\over 4}(E_{\rm C}^2-E_{\rm J}^2-\Delta^2)\epsilon^2\nonumber\\
&+&\textstyle{1\over 4}E_{\rm C}E_{\rm J}^2\epsilon-\textstyle{1\over 16}E_{\rm C}^2E_{\rm J}^2=0~.
\label{eqn:roots}
\end{eqnarray}
The population difference between the left and right wells is
obtained from the eigenenergies via
\begin{eqnarray}
N&=& \frac{\Delta}{(E_{\rm C}-2\epsilon)}~.
\end{eqnarray}
An important feature of equation (\ref{eqn:roots}) is that there are
four real roots when the coefficient of the quadratic term
$E_{\rm C}^2-E_{\rm J}^2-\Delta^2$ is positive and only two when it is
negative. For a symmetric double-well ($\Delta=0$), the
additional roots appear at the critical point where the
self-interaction energy satisfies $E_{\rm C}=E_{\rm J}$. For $E_{\rm C}>E_{\rm J}$, the
eigenenergies of the stationary states are $\pm
E_{\rm J}/2$ and $E_{\rm C}/2$, with the latter being doubly degenerate. The
additional roots disappear if the displacement of the barrier is such
that $\vert\Delta\vert=\vert2\alpha x_0\vert> \sqrt{E_{\rm C}^2-E_{\rm J}^2}$.
%The ground and first excited state are, respectively, even and odd
%states while the degenerate excited states are localised primarily in
%either the left or the right well.

To test the applicability of the two-state model, we have 
compared the eigenenergies with those determined from the numerical
solution of the GP equation.
The energy curves are parametrised by three numbers: the splitting between
the two lower levels, which is equal to $E_{\rm J}$;
the energy of the self-trapping states, $E_{\rm C}/2$; and an energy gradient, $\alpha$.
In a two-state model, $\alpha$ is both the energy gradient of
the self-trapping state, and that of the ground state 
for $x>x_{\rm c}$. For small $g$ and high
barriers, such that $E_{\rm C}\gg E_{\rm J}$, $\alpha\sim-E_{\rm C}/2x_{\rm c}$.

\begin{figure}[hbt]
\centering
\epsfig{file=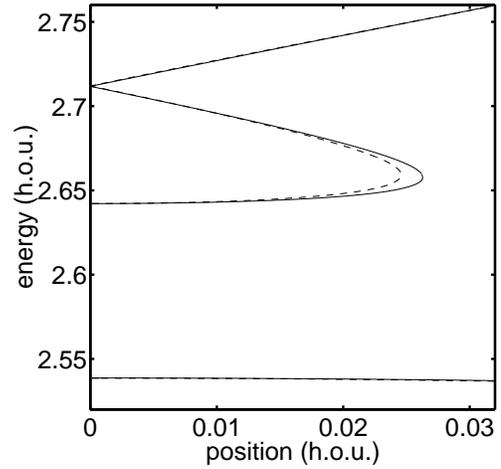,width=6.5cm,angle=0,clip=,bbllx=70,bblly=170,bburx=495,bbury=585}
\caption{Comparison between the exact (solid) and the two-state model (dashed)  
eigenenergies as a function of the barrier position for $h=4$ and $g=0.5$.}
\label{fig:model_e}
\end{figure}

Fig.~\ref{fig:model_e} shows a comparison between the exact eigenenergies
and the model curves for $h=4$, $g=0.5$. The values of $E_{\rm C}$, $E_{\rm J}$, and $\alpha$ are taken from the exact solutions.
The energy gradient is matched to that of the self-trapping states at $x_0=0$ ($\alpha=3.187$). This value gives the best agreement
when comparing population dynamics (see Section~\ref{sec:dynamics}). 
However, due to the slight 
curvature of the eigenenergy curves a smaller value ($\alpha=2.663$) gives 
a better fit to the triangular structure.
The agreement between the two-state model and the exact 
eigenenergies becomes less good as the influence of 
higher-lying states increases.
For example, the model does not predict the almost
flat position dependence of the upper levels in $e_1$ for
$h=4$, $g=5$, see Fig.~\ref{fig:eh4c5}.

\begin{figure}[hbt]
\centering
\epsfig{file=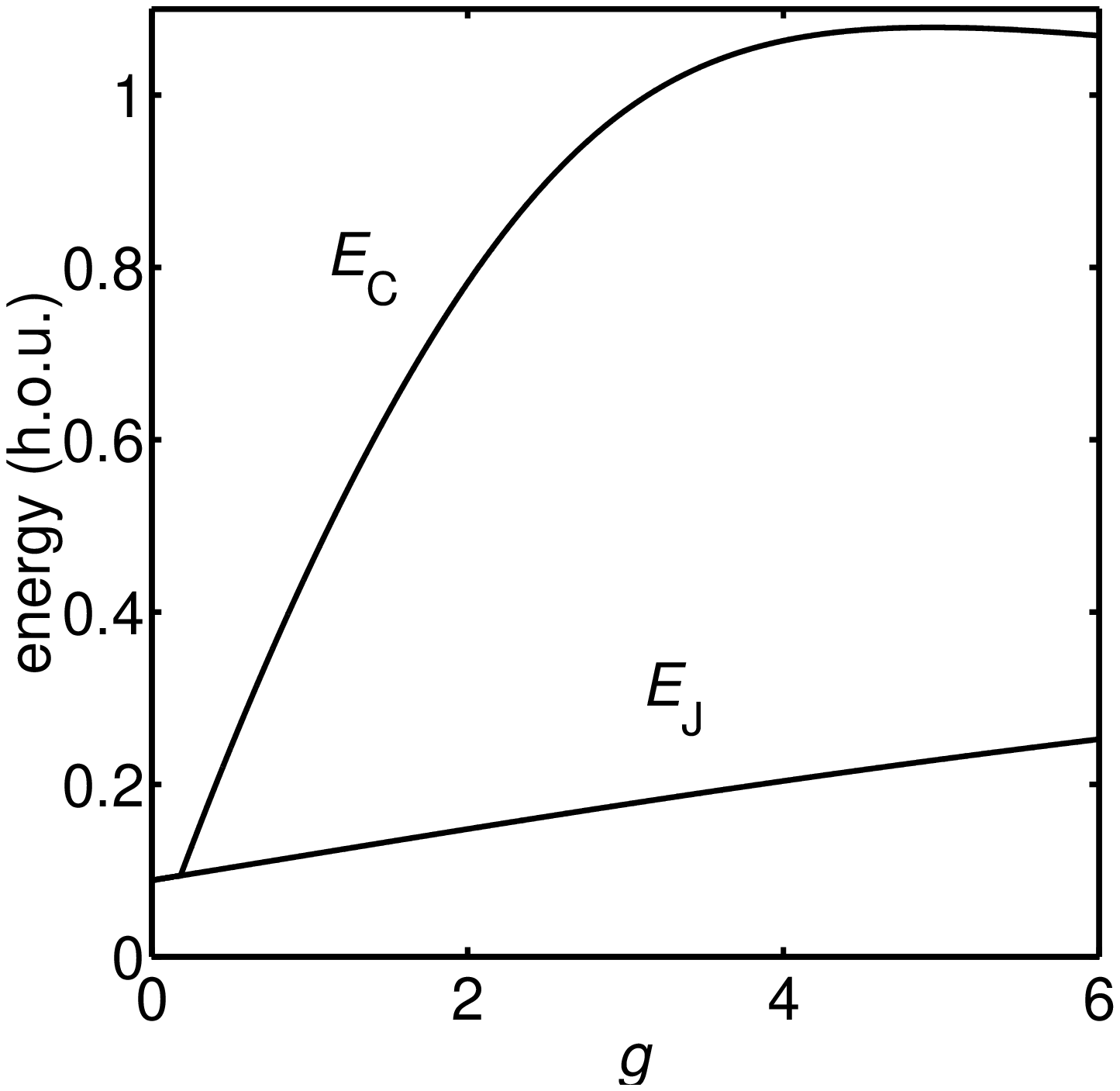,width=4.25cm,angle=0,clip=,bbllx=65,bblly=165,bburx=505,bbury=590}
\epsfig{file=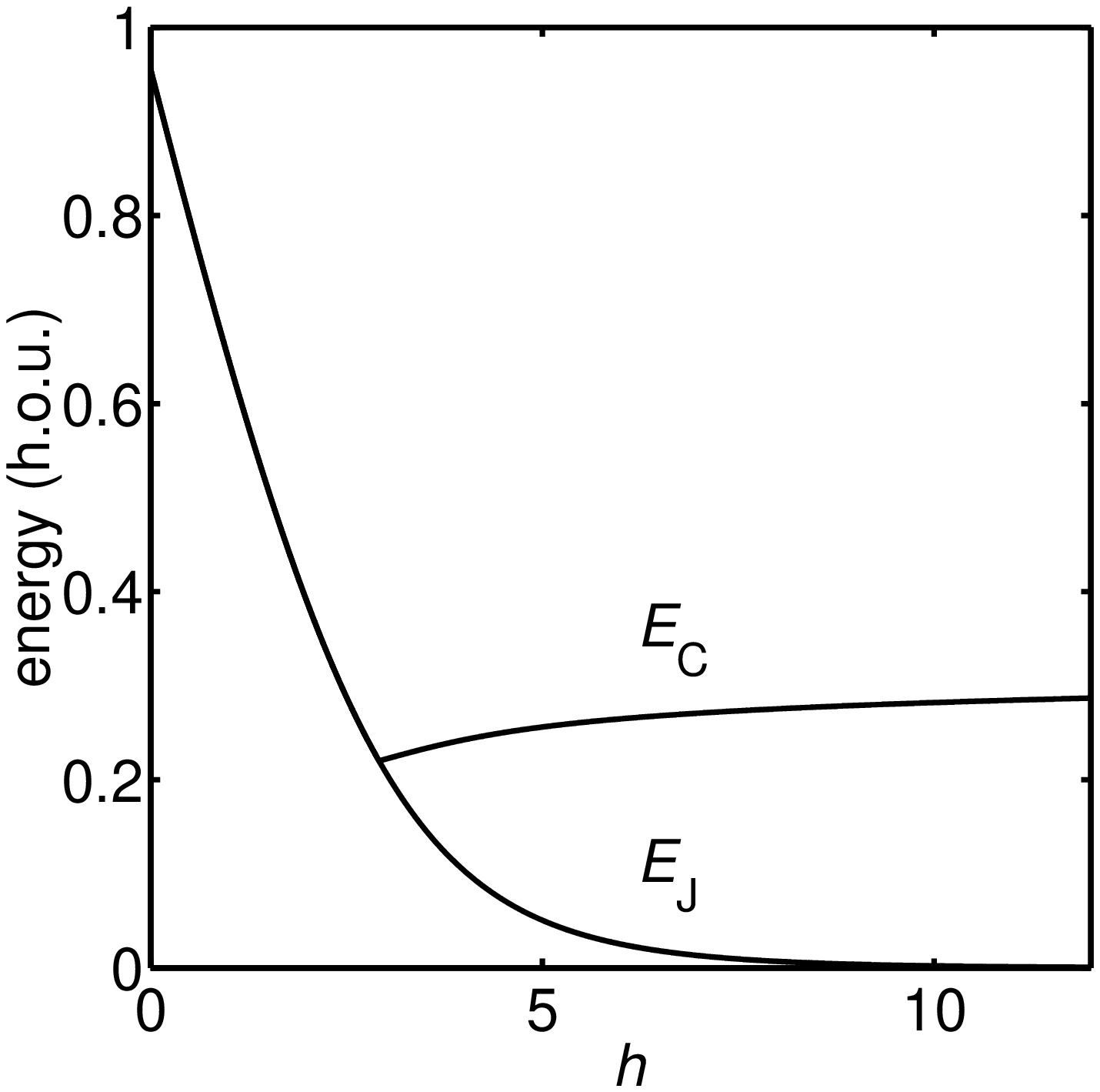,width=4.25cm,angle=0,clip=,bbllx=65,bblly=165,bburx=505,bbury=590}
\caption{The dependence of the self-interaction energy, $E_{\rm C}$, and the 
Josephson energy, $E_{\rm J}$, on the self-interaction parameter, $g$
(with $h=4$), and the barrier height, $h$ (with $g=0.5$). 
}
\label{fig:edep}
\end{figure}

To apply the two-state model one needs to known the value of
$E_{\rm J}$, $E_{\rm C}$ and $\alpha$ for any particular barrier
height or non-linearity. In Fig.~\ref{fig:edep}
we show how the energy splittings vary with the non-linearity 
and the barrier height. The critical point, 
$E_{\rm C}=E_{\rm J}$, appears as either a 
critical non-linearity or a critical height depending
on which parameter is varied. Note that for large non-linearity, 
the existence of the second excited state, $e_2$, effectively
puts an upper limit on the value of $E_{\rm C}$. 

\begin{figure}[hbt]
\centering
\epsfig{file=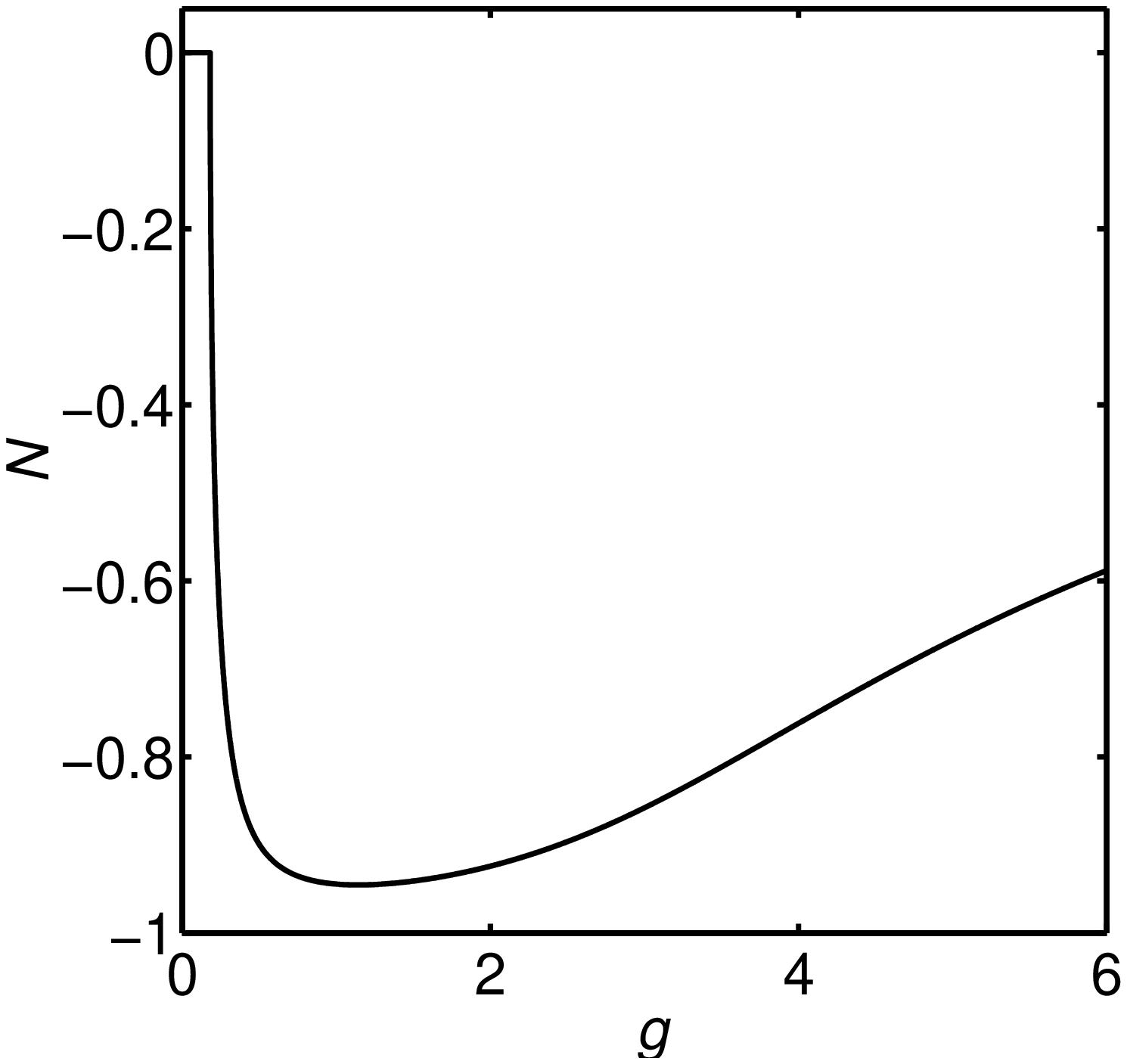,width=4.25cm,angle=0,clip=,bbllx=70,bblly=175,bburx=500,bbury=585}
\epsfig{file=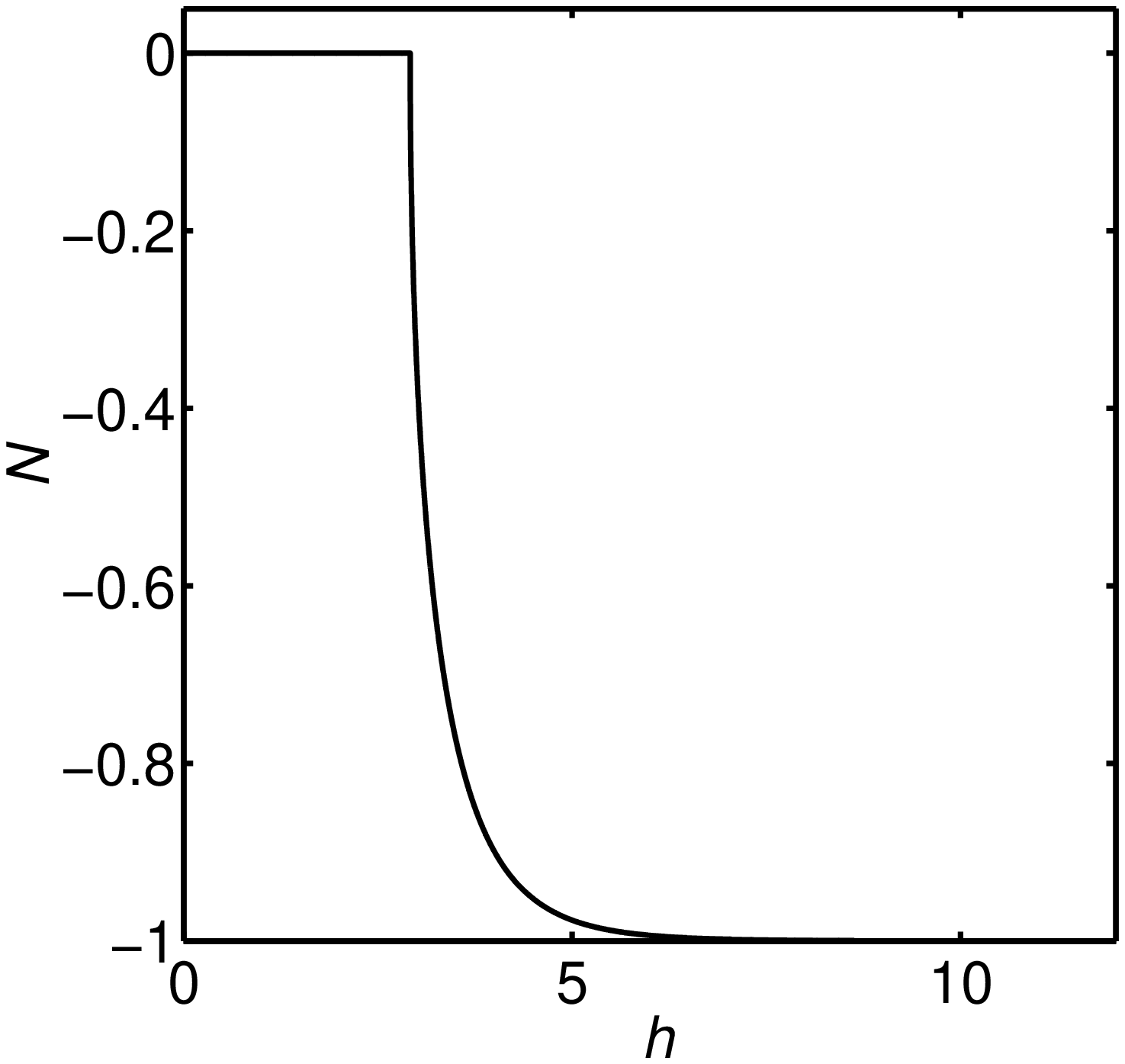,width=4.25cm,angle=0,clip=,bbllx=70,bblly=175,bburx=500,bbury=585}
\caption{The population asymmetry of the self-trapping state as a function of 
the self-interaction parameter, $g$ (with $h=4$),  and the barrier height, $h$
(with $g=0.5$).}
\label{fig:ndep}
\end{figure}

In Fig.~\ref{fig:ndep}, we show the left-right population difference of the self-trapping
state. A population asymmetry
appears at the critical point, $E_{\rm C}=E_{\rm J}$, corresponding to the
critical barrier height, $h=2.916$ for $g=0.5$. The asymmetry
increases with increasing barrier height. As a function of the 
non-linearity, the self-trapping population
first increases, then saturates, and finally decreases
at large $g$ due to the influence of $e_2$.

\section{Josephson dynamics}
\label{sec:dynamics}

We now apply the eigenstate picture to analyse
the population dynamics when the barrier is moved uniformly through the
condensate at velocity $v$. The time-dependent GP
equation is integrated using a Crank-Nicholson algorithm and the
time evolution analysed in terms of transitions between eigenstates.
A similar approach was used previously to study vortex nucleation
\cite{wini00}.

\begin{figure}[hbt]
\centering
\epsfig{file=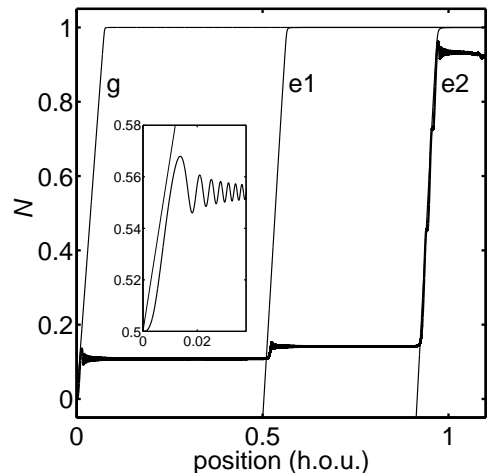,width=6.5cm,angle=0,clip=,bbllx=65,bblly=180,bburx=570,bbury=680}
\caption{
The population asymmetries for eigenstates $g$ , $e_1$ and $e_2$
 as a function of the barrier position with $h=12$ and
$g=0.5$. The population difference as the barrier is
moved from the centre towards the right at speeds of 4$\times10^{-5}$ 
is plotted as a thick black line. An expanded view of the evolution for 
short times is shown inset.}
\label{fig:nh12}
\end{figure}

In Fig.~\ref{fig:nh12} we show both the eigenstate population 
differences and the calculated evolution of the 
population difference as the barrier is moved from
the canter towards the right at a speed $v=4\times10^{-5}$ (bold line). 
This example is similar to the case
studied in \cite{giov00}. The eigenstate population 
differences appear as straight diagonal lines.  
For low velocities, the evolution is adiabatic, and
the time-dependent solution follows the ground state population difference
curve resulting in a dc tunnelling current. Above a critical velocity, there
is a transition to a superposition of ground and excited states such that
the population difference remains constant, apart from a decaying oscillatory
component (see inset). Subsequently, the excited state component 
encounters a level crossing with higher-lying states which 
leads to further `steps' in the population 
difference or resonances in the tunnelling current. 
For the example of $v=4\times10^{-5}$ shown in Fig.~\ref{fig:nh12}, the
speed is just below the critical velocity for the transition 
from $e_2$ to $e_3$ leading to a large tunnelling current as 
the object moves past $x_0\sim0.9$.
 
\subsection{Critical velocity}

By defining $\theta(t)=\theta_1(t)-\theta_2(t)$, and rearranging
equations (\ref{eqn:coeff}--\ref{eqn:NLHam}), one obtains the coupled
equations
\begin{equation}
\dot{N}=E_{\rm J}\sqrt{1-N^2}\sin\theta~,
\label{eq:ndot}
\end{equation}
\begin{equation}
\dot{\theta}=\Delta-NE_{\rm C}-\frac{E_{\rm J}N}{\sqrt{1-N^2}}\cos\theta~.
\label{eq:thetadot}
\end{equation}
Differentiating again and substituting for $\dot{N}$ and $\dot{\theta}$ gives
\begin{equation}
\ddot{\theta}=\dot{\Delta}-\frac{E_{\rm J}(E_{\rm C}-N\Delta)}{\sqrt{1-N^2}}\sin\theta
-\frac{E_{\rm J}^2(1+N^2)}{2(1-N^2)}\sin2\theta~.
\label{eq:thetaddot}
\end{equation}
If $N$ is roughly constant, this equation 
may be re-written as $\ddot{\theta}=-\partial_\theta U$, where
$U(\theta)$ describes a
`tilted-washboard' potential. The critical velocity is
determined by setting the minimum gradient of $U(\theta)$ equal to zero. 
For $N\approx 0$ at $t=0$ and using $\Delta=-2\alpha x_0=-2\alpha vt$, one finds 
\begin{equation}
v_{\rm c}=-\frac{E_{\rm J}E_{\rm C}}{2\alpha }-\frac{E_{\rm J}^2}{4\alpha}~.
\end{equation} 
For $E_{\rm C}\gg E_{\rm J}$, one can use 
the linear approximation, $\alpha=-E_{\rm C}/2x_{\rm c}$, 
where $x_{\rm c}$ is the critical displacement,
which yields $v_{\rm c} \sim E_{\rm J}x_{\rm c}$.
Taking the values from Fig.~\ref{fig:eh12}, $E_{\rm J}=7\times 10^{-4}$ and $x_{\rm c}=0.07$, gives $v_{\rm c}\sim5\times 10^{-5}$.
According to equations (\ref{eq:ndot}) and (\ref{eq:thetadot}) 
the maximum population difference occurs at 
approximately $v_{\rm c}/\sqrt{2}= 3.4\times10^{-5}$, which is in excellent
agreement with the value determined by numerical integration
of the GP equation. 
 
\subsection{Asymmetric initial condition}

\begin{figure}[hbt]
\centering
\epsfig{file=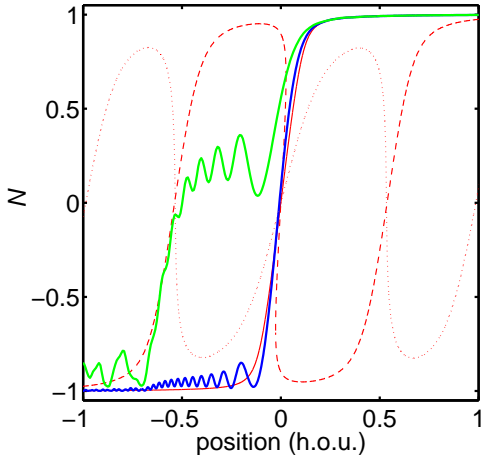,width=6.5cm,angle=0,clip=,bbllx=65,bblly=170,bburx=500,bbury=585}
\caption{The population difference as the barrier is moved from
$x_0=1$ towards the left at a constant speed of 0.007 (black) and 0.015
(grey) for $h=4$ and $g=0.5$.  
The corresponding population differences for eigenstates $g$, $e_1$, and $e_2$
are indicated by the solid, dashed and dotted lines.
For $e_2$, the gradient of $N$ against $x_0$ is
always positive, whereas for $e_1$ the gradient
becomes negative in the vicinity of the level crossing. Recalling that
when the barrier is moving the tunnelling current is proportional to this gradient, 
it follows that the appearance of the triangular structure in 
$e_1$ is associated with an infinite tunnelling current and a complete 
breakdown of adiabatic evolution.
}
\label{fig:nh4cp5}
\end{figure}

As stated above, the small critical velocity for high barriers (typically 
less than a micron per second) makes experimental verification
challenging. Although higher critical
velocities are obtained for lower barriers, the transition
between the dc and ac regime becomes less sharp. This problem
can be partially circumvented by noting that
one can induce larger population changes by starting with an asymmetric
well and moving the barrier back through the origin.  This is
illustrated in Fig.~\ref{fig:nh4cp5} with
a lower barrier height, $h=4$. Again
for low barrier speeds the population follows
the ground state distribution, whereas for faster speeds
a transition to an oscillatory current is observed.
The critical velocity is a factor of 200
times larger than for $h=12$. 

\begin{figure}[hbt]
\centering
\epsfig{file=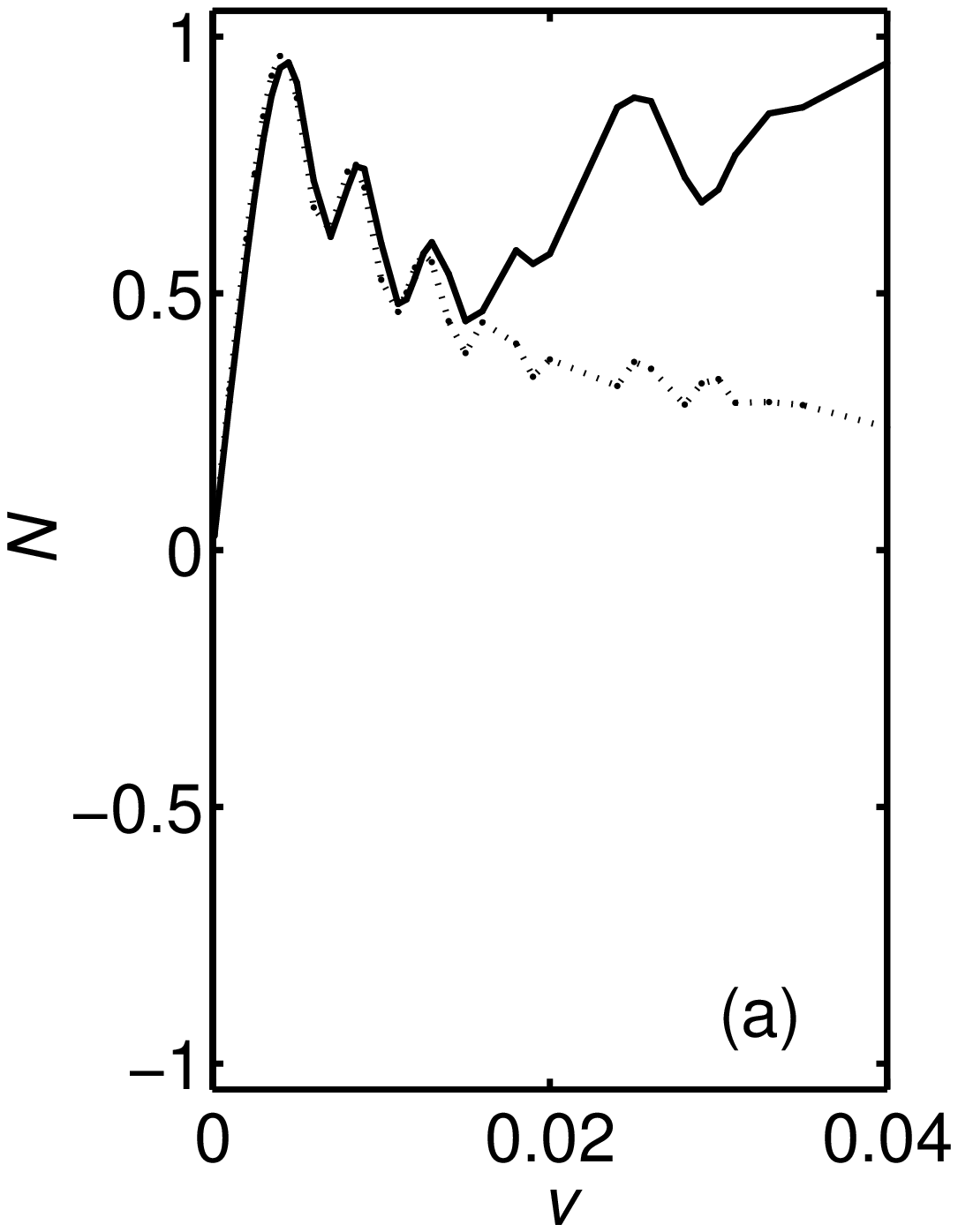,width=4.25cm,angle=0,clip=,bbllx=135,bblly=180,bburx=455,bbury=580}
\epsfig{file=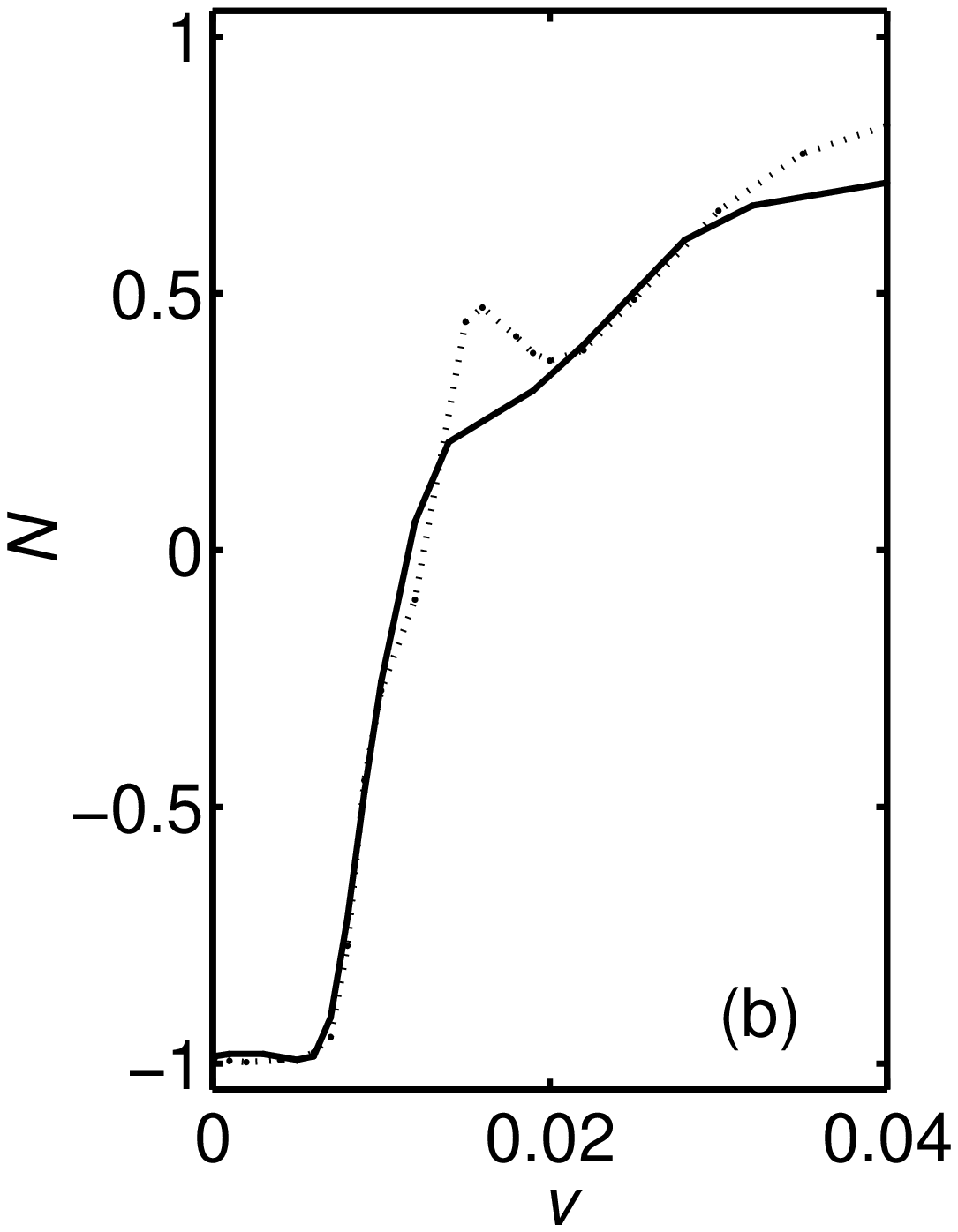,width=4.25cm,angle=0,clip=,bbllx=155,bblly=180,bburx=475,bbury=580}
\caption{Comparison of the Josephson tunnelling with (a) a symmetric
and (b) an asymmetric initial condition. The population difference as a function of
the barrier speed $v$ obtained from the GP equation (solid line)
and the two-state model (dashed line) for $h=4$ and $g=0.5$. 
In (a) the barrier is moved from the 
centre outwards for a time $t=30$. For $v>0.015$ the 
second level crossing is reached leading to
a departure between the GP simulation
and the prediction of the two-state model. In (b) the barrier is moved from
$x_0=0.95$ and stopped at $x_0=-0.3$.
The parameters $E_{\rm C}$, $E_{\rm J}$,
and $\alpha$ for the model are taken from the 
GP eigenenergies. }
\label{fig:nv}
\end{figure}

For $x_0<-0.6$ the population difference follows that of the excited state, $e_1$,
therefore to observe a large population difference, it is important to
stop the motion before the barrier reaches $x_0 \sim -0.4$.
This restricts the evolution to short times or low velocities. 
In Fig.~\ref{fig:nv} we show the population difference as a function of
the barrier speed for both a symmetric and 
asymmetric starting condition. 
Fig.~\ref{fig:nv}(a) shows that the two-state model can no longer
predict the correct result if the second level crossing is 
reached, which for $t=30$ occurs when $v>0.015$.
 
Comparison between Fig.~\ref{fig:nv}(a) and (b) illustrates that a
larger population difference and a better demarcation of
the critical velocity is obtained by moving
the barrier from the edge of the condensate inwards (asymmetric initial
condition). The numerical results
indicate that the critical velocity for this case 
is similar to that for the symmetric initial state. 
However, it is not straightforward to predict
the critical velocity for the asymmetric initial condition because
the relative importance of the coefficients in equation (\ref{eq:thetaddot})
depends on the instantaneous value of $N(t)$. 
For $h=12$, $g=0.5$, the critical velocity 
is a factor of three smaller than for the symmetric initial state.  

\begin{figure}[hbt]
\centering
\epsfig{file=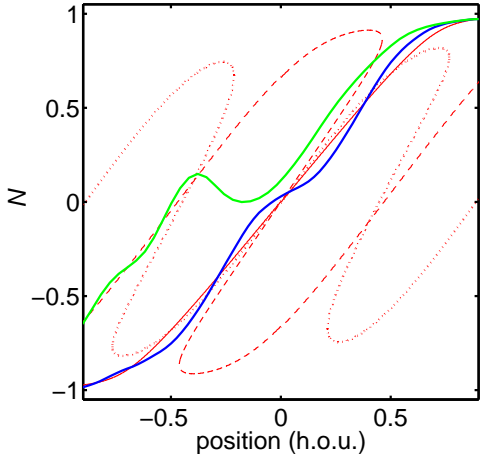,width=6.5cm,angle=0,clip=,bbllx=65,bblly=170,bburx=500,bbury=585}
\caption{The population difference as
the barrier is moved from position 0.9 towards the left at speed of
0.05 (black) and 0.08 (grey) for $h=4$ and $g=5$.  
The corresponding population asymmetries for
eigenstates $g$, $e_1$, and $e_2$ are indicated by the solid, dashed and dotted
lines.}
\label{fig:nh4c5}
\end{figure}

Finally, we consider the interesting case where the influence
of higher-lying states severely limits the applicability of 
the two-state model. In Fig.~\ref{fig:nh4c5} we show
the population change as a barrier with height $h=4$ 
is moved through the centre of a condensate with $g=5$.  
For these parameters the critical velocity is sufficiently 
high that the level crossing to a higher-lying state is 
reached before the transition to the ac regime is 
completed. However, a large population difference for different
barrier velocities can still be observed 
if the motion is halted when the barrier reaches $x_0\sim -0.5$.
 
The large critical velocity, $v_{\rm c}\sim 0.08$, offers the best
potential for the experimental observation of Josephson-like
tunnelling resonances.  For example, taking 
a typical sodium condensate 
with a trap frequency of $20$~Hz, the black and grey curves in 
Fig.~\ref{fig:nh4c5} correspond to a
barrier formed by a blue detuned laser sheet with waist 3 microns,
moved by 5 microns in 1.5~s and 1~s, respectively.
%An attractive system for exploring the Josephson
%effect is a condensate where the interactions can be tuned
%via a Feshbach resonance \cite{corn00} thereby allowing complete control
%over both the non-linearity and the barrier height. 
For these parameters, a model involving higher-lying excited 
states is essential to give
a clear picture of the Josephson population dynamics.

To restate the richness of the double-well system, it is worth noting
that as one increases the non-linearity further, the transition from
the ground to first excited state corresponds to the formation of a
soliton. Consequently by varying only two parameters, the barrier
height and the interaction strength, for example using a 
Feshbach resonance \cite{corn00}, one can explore the complete
parameter space between Josephson tunnelling and soliton formation.

\section{Conclusions}

In summary, we have studied the eigenstates of a dilute Bose-Einstein
condensates in an asymmetric double-well potential.
We have shown that in the regime of
interest for experiments, i.e., with a large non-linearity and a low 
barrier, a two-state description
is insufficient to describe the tunnelling dynamics.
By determining the influence of higher-lying states we have demonstrated
that large population differences can be observed if
the barrier motion is halted before the upper tunnelling resonance
is reached.

\acknowledgements
We thank Andrew MacDonald whose preliminary study stimulated our 
interest in this problem. 
We would also like to thank the University of Durham and the Engineering and
Physical Sciences Research Council (EPSRC) for financial support.

\end{document}